\def\kms{km~s$^{-1}$}
\def\etal{et~al.~}
\documentstyle[11pt,aaspp4,flushrt]{article}

\tighten
\begin{document}

\title{Establishing the Connections Between Galaxies and Mg II
Absorbing Gas\altaffilmark{1}}
\pagestyle{empty}

\author{\sc Christopher W. Churchill\altaffilmark{2}}
\affil{Lick Observatory, University of California, Santa Cruz, CA 95064}
\altaffiltext{1}{To appear in {\it IAGUSP Workshop on Young Galaxies and
QSO Absorbers}, eds.~S.M.~Viegas, R.~Gruenwald, \& R.~de~Carvalho,
(PASP Conference Series)}
\altaffiltext{2}{Visiting Astronomer, Astronomy and Astrophysics
Department, Pennsylvania State University}

\begin{abstract}
HIRES/Keck spectra of Mg II $\lambda 2796$ absorption arising 
in the ``halos'' of 15 identified $0.4 < z < 0.9$ galaxies are
presented.
Comparison of the galaxy and absorbing gas properties reveal that the
spatial distribution of galactic/halo gas does not follow a smooth
galactocentric dependence. 
The kinematics of absorbing gas in $z\sim 1$ galaxies are not
suggestive of a {\it single}\/ systematic velocity field (i.e.~rotation
or radial flow) and show little dependence on the QSO--galaxy impact
parameter.
From the full HIRES dataset of 41 systems ($0.4 < z < 1.7$), strong
redshift evolution in the cloud--cloud velocity dispersion is
measured.
Direct evidence for turbulent or bulk motion in ``high velocity''
clouds is found by comparing Fe~II and Mg II Doppler parameters.
\end{abstract}

\section*{Introduction}
\pagestyle{myheadings}
\markboth{\sc C.W.~Churchill \hfill Galaxies and {\rm Mg} II Absorbing Gas~~}
         {\sc C.W.~Churchill \hfill Galaxies and {\rm Mg} II Absorbing Gas~~}

The study of Mg II absorption in QSO spectra provides a powerful tool
for studying the evolution of the kinematic and spatial distribution
of galactic/halo gas in galaxies from the epoch of the first QSOs.  
Ultimately, these studies will provide observational constraints on 
the mechanisms and frequency of events by which galaxies are
constructed, which in turn will provide an independent and direct
quantification of the development of $10^{12-13}$~M$_{\sun}$ structure
formation in the universe.

It has been a few short years since Petitjean \& Bergeron (1990)
undertook the first investigation of the kinematic and spatial
distribution of Mg II absorbing gas based upon the subcomponent
clustering within the absorption profiles.
Shortly thereafter, Bergeron \& Boiss\'e (1991) were the first to
established that Mg II absorption lines in QSO spectra arise in the
proximity of apparently normal $L^{\ast}$ field galaxies.
With the more recent sample of Mg II absorbing galaxies compiled by
Steidel, Dickinson, \& Persson (1994, hereafter SDP) and the advent of
the HIRES spectrograph (Vogt \etal 1994) on the Keck 10--m telescope,
one can now undertake a detailed study designed to establish how
galaxy and absorption properties correlate.
Such a study promises to provide the information necessary for
developing a more detailed picture of galaxy/halo gas substructures
and their dynamics (Lanzetta \& Bowen 1990, 1992; Charlton \&
Churchill 1996; Charlton \& Churchill, this volume).

This contribution is a partial description of my thesis work and 
of work that appears elsewhere [Churchill, Steidel, \& Vogt 1996
(hereafter CSV); Churchill, Vogt, \& Charlton 1996].

\section*{The Project}

The HIRES sample was selected on the basis that each absorption
system was associated with an imaged galaxy that was spectroscopically
confirmed to have the same redshift as seen in absorption (SDP).
For each galaxy, the rest $L_B$ and $L_K$ luminosities, rest $B-K$
colors, QSO--galaxy impact parameters, $D$, and redshifts, $z_{\rm
gal}$, are measured. 

\begin{figure}[bh]
\plotfiddle{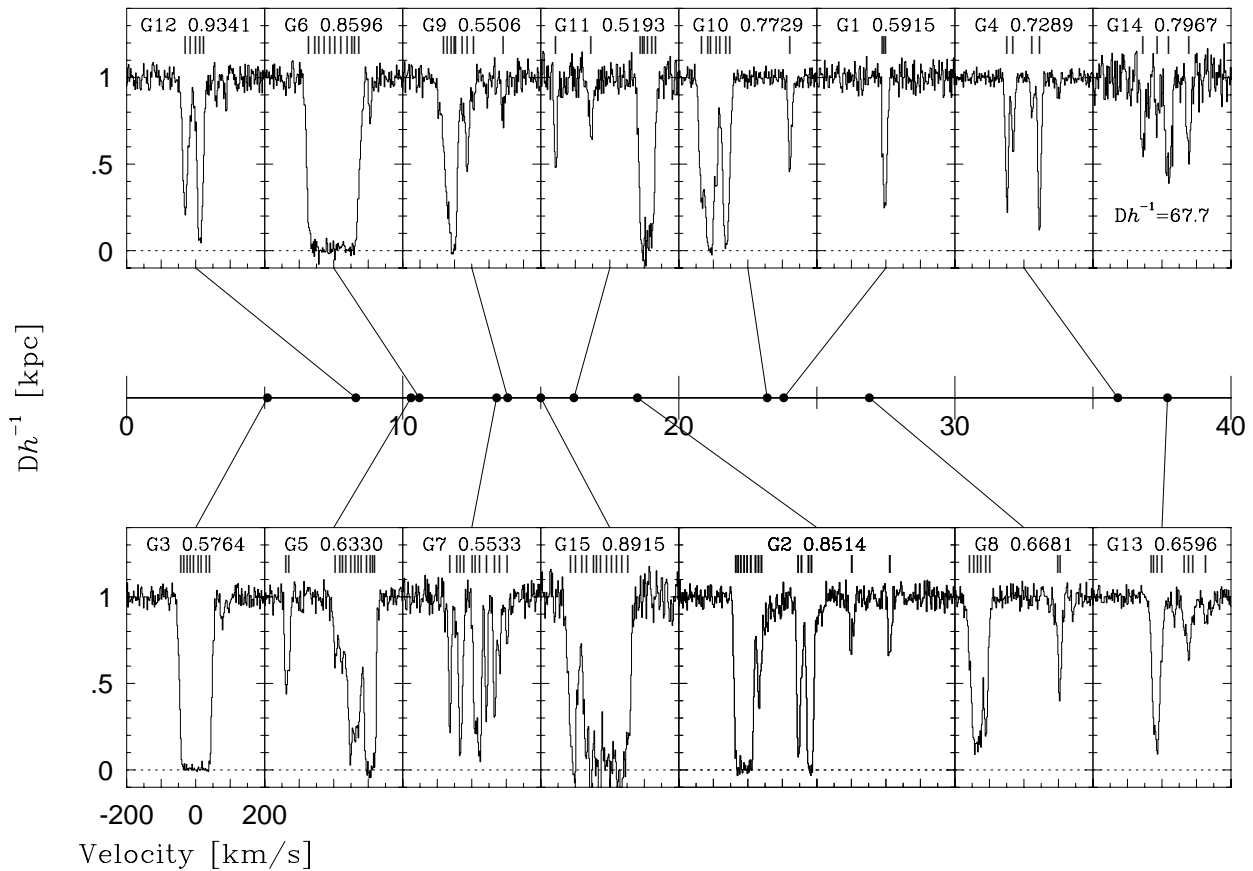}{4.55in}{0}{68.0}{68.0}{-265}{-35}
\caption
{
The HIRES/Keck absorption profiles of the Mg II $(\lambda 2796)$
transition in order of increasing impact parameter, $Dh^{-1}$~kpc
(taken from CSV).  The redshifts and galaxy IDs, as tabulated in Table 2
of CSV, are given.  The vertical ticks above the continuum of each
absorption feature mark the subcomponents used in the kinematic
analysis.  Each panel has a velocity spread of 400~{\kms}, except G2,
which has an 800~{\kms} spread.  Note that $D\sim 70$~kpc for G14.
}
\end{figure}

One motive for the study was to test the paradigm suggested by
Lanzetta \& Bowen in which intermediate redshift galactic
halos are roughly identical, having cloud spatial number density
distributions $\propto r^{1-2}_{\rm gal}$ and systematic rotational or
radial flow kinematics.
The primary test of this scenario is the prediction that the {\it
observed}\/ differences in the absorption properties from one system
to another are predominantly due to the QSO--galaxy impact parameter.
In Figure 1, the Mg II $(\lambda 2796)$ transitions are shown in order
of increasing QSO--galaxy impact parameter ($q_0 = 0.05$ and $h =
H_0/100$~{\kms}~Mpc$^{-1}$).  
The gas properties have been parameterized from the absorption
profiles using Voigt profile fitting (cf.~Carswell \etal 1991), which
yields the number of ``clouds'' and their individual column
densities, Doppler $b$ parameters, and velocities.
As described by CSV and given in Table 2 of their work, the gas
kinematics of each system have been characterized by
(1) the number of clouds, 
(2) the median absolute deviation of cloud velocities, 
(3) the number of median absolute deviations of the highest velocity
    cloud, 
(4) and the velocity asymmetry.
The absorption strengths are measured by the rest equivalent widths
and doublet ratios.

\section*{Results}

Spearman and Kendall non--parametric rank correlation tests were used
to ascertain if galaxy properties correlate with the absorption
strengths and kinematic indicators.  
The tests revealed that the null--hypothesis of no correlation is 
consistent with the data when the criterion of a greater than 97\%
confidence level is applied.
However, trends with large scatter are not ruled out (cf.~Steidel
1995, Fig.~3; CSV).
{\it Of primary significance is the fact that the QSO--galaxy impact
parameter does not provide the primary distinguishing factor by which
absorption properties can be characterized}.
The implication is that the spatial distribution and kinematics 
of absorbing gas surrounding intermediate redshift galaxies is not
roughly identical from galaxy to galaxy, even if the processes that
give rise to the gas are.

In Figure 2, results from profile fitting to the full HIRES dataset
are shown, where only 5$\sigma$ detections have been included.
As shown in the left--hand panels, the sample was sub--divided by the
median redshift, $\left< z \right> \sim 0.9$, and the cloud--cloud
velocity differences within each system were computed.
The histograms are the relative number of cloud--cloud pairs in each
20~{\kms} bin.  
The solid lines are fitted Gaussians and give the cloud--cloud
velocity dispersion of the absorbing gas.
The kinematic dispersion is seen to evolve with redshift, such that
$\sigma(\Delta v) \sim 140$~{\kms} at $\left< z \right> = 1.2$
decreases to $\sigma(\Delta v) \sim 60$~{\kms} at $\left< z \right> =
0.6$.
The number of clouds do not evolve with redshift, so one can conclude
that the observed evolution of large equivalent width systems
(Steidel \& Sargent 1992) is in fact due to a settling of the velocity
dispersion.
  
For all 41 systems, the distribution of Doppler $b$ parameters,
$b$(Mg II), is peaked at $\sim 4.9$~{\kms} and is consistent with a
Gaussian with $\sigma \sim 1.5$~{\kms} that is truncated below
3.0~{\kms}.
Using the atomic masses of Fe and Mg, the bulk/turbulent $b$ parameter
has been computed for {\it unblended}\/ clouds assuming that the
thermal and turbulent components can both be represented as Gaussians
($b_{\rm tot}^{2} = b^{2}_{\rm therm} + b^{2}_{\rm bulk}$, where
$b^{2}_{\rm therm} \propto m_{\rm ion}^{-1}$).
The typical Fe II and Mg II $b$ parameter uncertainties are
$\sigma (b_{\rm tot}) \sim 0.25$~{\kms}.
As a result of selecting unblended lines, these clouds all
happen to have velocities in excess of 100~{\kms} from the profile
optical depth weighted velocity zero point.
It is apparent that some fraction of these ``high velocity'' clouds
exhibit both a thermal broadening and a turbulent or bulk motion
component.

\begin{figure}[bh]
\plotfiddle{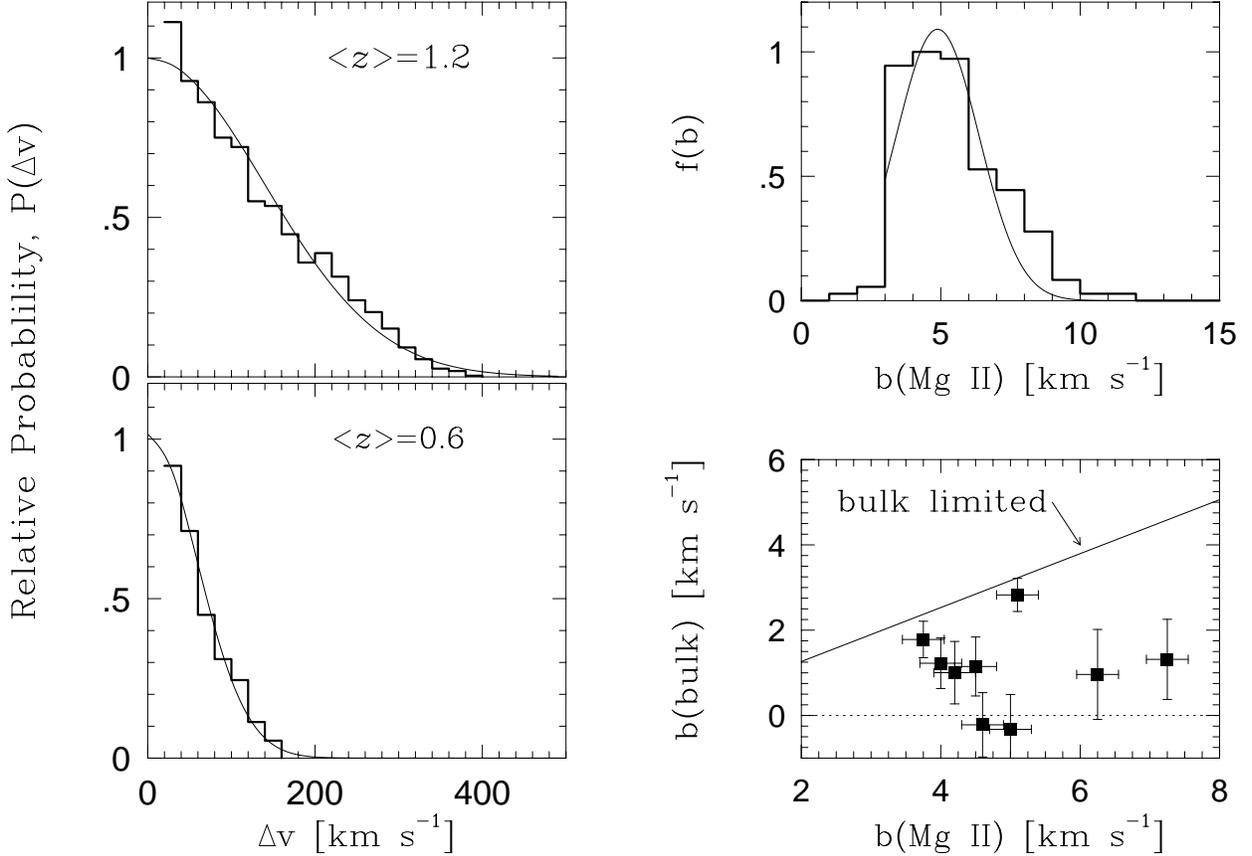}{4.55in}{0}{66.0}{66.0}{-250}{-35}
\caption
{
Results from profile fitting to the full HIRES dataset.
The left panels illustrate the redshift evolution of the gas
kinematics.  Shown are the normalized probabilities that any two
clouds in a given system have a line of sight velocity difference 
$\Delta v$.  The upper right panel illustrates the distribution of
Doppler $b$ parameters.  The lower right panel shows the degree to
which unblended ``high velocity'' clouds exhibit bulk or turbulent
internal motions.  See the text for details.
}
\end{figure}

Future plans include incorporating the galaxy morphologies and
orientations with respect to the QSO light path from Hubble Space
Telescope images obtained by Chuck Steidel.  
The new STIS should be ideal for obtaining the wider range of higher 
ionization species needed for a more complete picture of the kinematic,
chemical, and ionization conditions of early epoch galactic gas.

\newpage
I thank Jane Charlton, Chuck Steidel, and Steven Vogt for their
continued support, assistance, and pleasurable collaborations. 
This work supported in part by the California Space Institute, and
NASA (grant NAGW--3571).

\end{document}